\documentclass[12pt]{article}  

\usepackage{amsmath,amsfonts,amssymb}
\usepackage{graphicx}
\usepackage{float}
\usepackage[colorlinks=true, allcolors=blue]{hyperref}
\usepackage{makecell}

\providecommand{\keywords}[1]
{
  \small	
  \textbf{\textit{Keywords---}} #1
}

\title{Optical design options for Pollux: UV spectropolarimeter project for the Habitable Worlds Observatory}
\author{Eduard Muslimov$^{a,c}$,Coralie Neiner$^{b}$,\\
Jean-Claude Bouret$^{c}$ \\
\small a--Department of Physics, University of Oxford, \\
\small Keble Rd, OX14 3RH Oxford, UK\\
\small b -- LESIA, Paris Observatory, PSL Research University,\\
\small CNRS, Sorbonne Université, Université  Paris Cité,\\
\small 5 place Jules Janssen, 92195 Meudon, France\\
\small c--Aix Marseille Univ, CNRS, CNES, LAM, Marseille, France\\
}

\begin{document} 
\maketitle{}

\begin{abstract}
Pollux is a project of UV spectropolarimeter proposed as a European contribution for the planned NASA-led Habitable Worlds Observatory. In the present study we consider design options for four spectral channels. The two main channels operate in the range of 118-472 nm. We consider a few design options depending on the hosting telescope size and the approach to correct the aberrations in the camera part and show that the resolving power of R $>$ 90'000 is reachable. In addition, we study 2 other channels: a visible and near infrared spectropolarimeter, which could reach up to 1050 nm or 1800 nm depending on the detector choice, and a far UV channel operating in the range of 100-120 nm. We also provide two design options with different resolution and main disperser type for these channels. 
\end{abstract}


\section{INTRODUCTION}
\label{sec:intro}  

In 2020 a review of a few proposals for the next generation flagships space observatories was conducted for the decadal survey from the U.S. National Academies of Sciences, Engineering, and Medicine\cite{NAP26141}. 
Based on the recommendations from this survey NASA has started the Habitable Worlds Observatory (HWO) project. Currently it is at an early development stage, so the full list of the science goals, top-level observatory architecture and the R\&D plan are still in progress \cite{Burns2024How}. However, the HWO project will rely on the heritage of the LUVOIR \cite{Roberge2021BAAS} and HabEx \cite{Martin8742208} proposals to a large extent. 

In this context we consider the redesign of the Pollux instrument proposal\cite{Muslimov10.1117/12.2310133} made for the LUVOIR-A observatory and its adjustment for the science goals and expected technical specifications of HWO. Initially Pollux had been conceived as a high-resolution ultraviolet (UV) echelle spectropolarimeter. However, as the entire HWO telescope project is at early stages of development with a few competing possible architectures, we should not restrict ourselves to a single design solution. Instead, in the present work we try to explore the landscape of possible solutions and highlight the main factors limiting the performance and feasibility in each case. 

Below we consider possible options for the instrument channels one-by-one. Each time we show the key technical decisions, which have to be taken and the corresponding trade-offs. We have to assume some inputs and target metrics regarding the telescope and the instrument. We use the following coarse estimates, which may not be enough to get precise quantitative answers, but should be suitable for a comparative study:
\begin{itemize}
    \item The telescope may have a primary mirror diameter of 6 or 8 m (see Lee D. Feinberg et al. paper at this conference).
    \item We assume $f/\#=26$ based on the LUVOIR-B architecture\cite{Corsetti10.1117/12.2530695}.
    \item Based on the science cases list compiled for Pollux for LUVOIR, \cite{Bouret10.1117/12.2312621} we set the required shortest wavelength equal to 118.5 nm to cover a broaden Lyman $\alpha$ line (with a 100 nm as a goal). 
\item Similarly, we assume that the longest wavelength	should be at least 390 nm with a goal higher than 400 nm.
\item We consider extension towards the visible (VIS) and near infrared (NIR) domain
at least up to 1200 nm as there are some analytical lines and bands significant for exoplanets characterisation in that range \cite{TENNYSON2024109083}, with a goal at 1800 nm.
\item The target spectral resolving power should be $R\geq90'000$ (with a top goal of 120'000) for the shortest wavelengths, although for the longer wavelengths it can be mitigated to $R\geq60'000$.
\item We target long spectral orders $\geq4 nm$ (with a goal of 5 nm) to avoid order stitching, which decreases the precision.
\item The slit/pinhole image should be Nyquist-sampled with $\geq2.5 pix$ ($\geq2.7 pix$ goal).
\item The instrument should provide circular+linear (QUV) polarization measurements, preferably with a possibility to switch to a pure spectral mode to gain in sensitivity for some science cases. 
\end{itemize}

\section{Medium and near UV: core instrument driven by the telescope architecture}
\label{sec:muvnuv}  

The very definition of the Pollux function, i.e. performing high-resolution spectropolarimetry at UV wavelengths unreachable with ground-based instrumentation, almost inevitably leads to a solution based on an echelle spectrograph with a dedicated polarimeter. Moreover, it has been shown that it is practically impossible to cover the entire range of interest in a single channel keeping the components parameters in any reasonable margins. Therefore, the core instrument represents two channels - near and medium UV (hereafter referred to as NUV and MUV), split by a dichroic filter. 

We have considered a few options for the sub-bands and found that taking into account the science goals, available materials and sensors, it would be just natural to split the UV domain into two octaves starting from the Lyman $\alpha$ line. Adding some margin to cover it securely, we obtain the sub-ranges of 118-236 nm and 236-472 nm for the MUV and NUV, respectively. This approach also makes the entire designs of the two channels geometrically almost identical, where only the gratings spatial frequencies and the diffraction orders numbers should be scaled accordingly.

\begin{figure} [!ht]
   \begin{center}
   \begin{tabular}{c} 
   \includegraphics[width=0.9\textwidth]{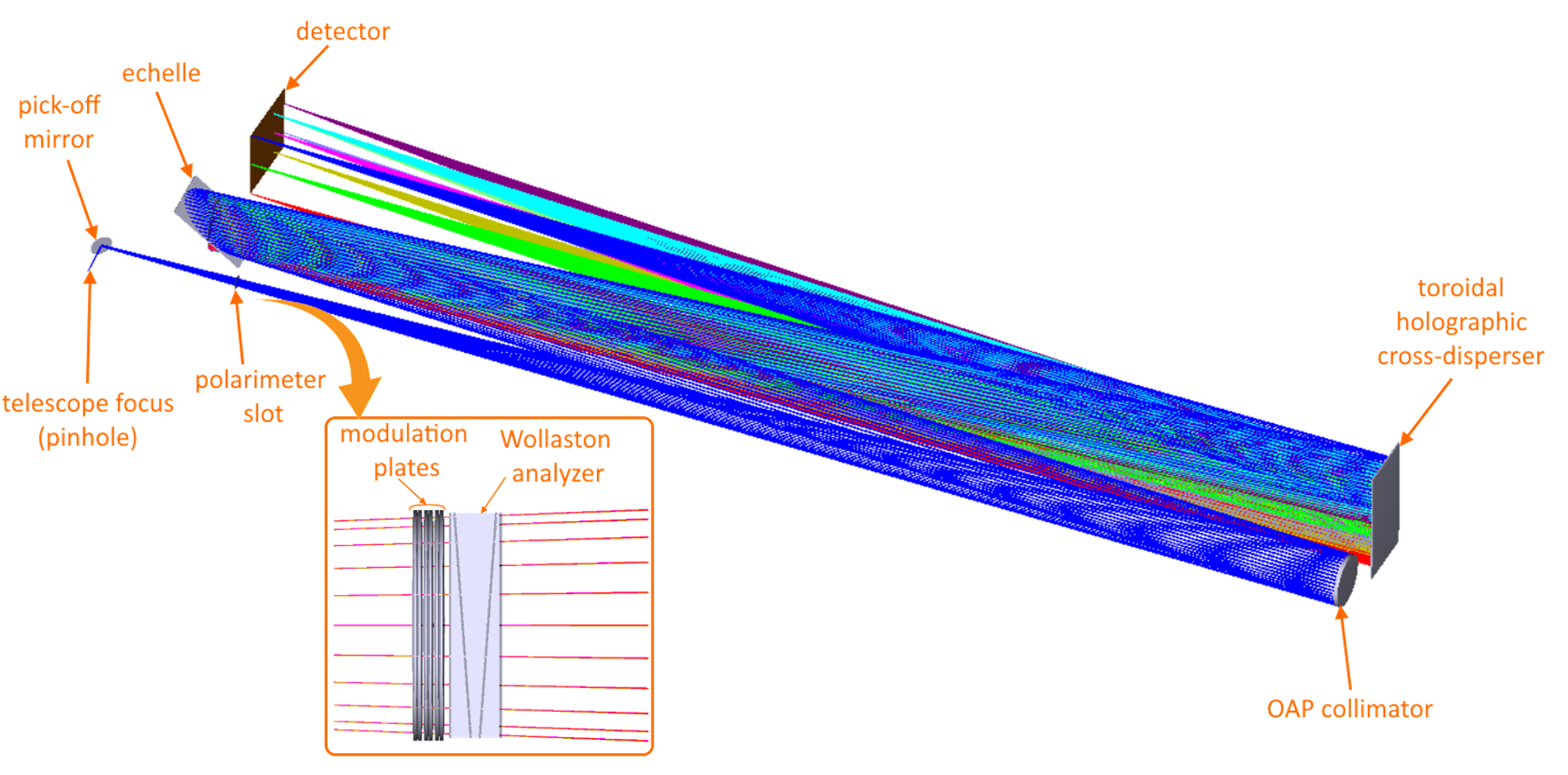}
   \end{tabular}
   \end{center}
   \caption[muv] 
   { \label{fig:muv} 
General view of the medium and near UV channels optical design. }
   \end{figure}

We assume that the instrument entrance represent a pinhole as we do not target any spatial resolution or covering an extended field of view. Its diameter should cover the Airy disk of the point-spread function (PSF) formed by the telescope with a $\approx 5 \mu m$ for the beam jitter \cite{Corsetti10.1117/12.2530695}, which gives $32.5 \mu m$ at the telescope focal plane, i.e. the instrument entrance. The collimator may be a single off-axis parabolic mirror, as it can provide a perfect collimated beam for a point source with a single reflection. The focal length is restricted by the envelope volume available for the instrument, so essentially, by the primary mirror size.  The main disperser for the specified target resolving power must be an echelle grating. Its parameters can be defined by  analytical equations\cite{eversberg2014spectroscopic}.  For the cross-disperser and camera we propose the same approach as was demonstrated earlier for Pollux@LUVOIR \cite{Muslimov10.1117/12.2310133}, i.e. using a concave grating performing both of the functions. Fortunately,  with a smaller telescope and reduced target resolution the angular field to cover and the dispersion of the incoming beam for this cross-disperser grating become smaller. Therefore, it is easier to reach the required aberration correction. For the spectropolarimeter operating in the chosen range we can apply the known solution\cite{Pertenais10.1117/12.2296215} based on $MgF_2$ plates with a tailored optical path difference and a Wollaston prism analyzer made of the same material. This unit can be retractable. Finally, here we assume that an EMCCD image sensor will be used for the given domain \cite{Nikzad_s16060927} as this was the choice for Pollux@LUVOIR, but CMOS will also be considered in the future as the optimal solution. General view of the optical design based on the elements and decisions listed above is shown in Fig.~\ref{fig:muv}.

At the very top level the complexity and reachable performance of this design is defined by the parameters of the hosting telescope. As it was mentioned before, it may change from 8 to 6 m and this directly affects the theoretical limit of the echelle spectrograph resolution.  Below in Table~\ref{tab:corePar} we compare the main design parameters obtained for each of the cases. In general, the parameters found for the 8m version look more feasible and safe in terms of the technology, instrument packaging and margins. In particular, the instrument's relative size is smaller in comparison with the primary mirror diameter and  the maximum gratings frequency is smaller.

 \begin{table}[!ht]
\caption{Key parameters of the MUV and NUV design options.} 
\label{tab:corePar}
\begin{center}       
\begin{tabular}{|l|c|c|c|c|c|c|c|} 
\hline
Parameter & \multicolumn{2}{|c|}{8m telescope} & \multicolumn{2}{|c|}{6m telescope}\\
\hline
 	&  NUV	& MUV &  NUV	& MUV \\
  \hline
Min wavelength, nm	&    236 &	118 &    236 &	118 \\
\hline
Max wavelength, nm	&472&	236&472&	236\\
\hline
Sampling, pix & \multicolumn{2}{|c|}{2.55}& \multicolumn{2}{|c|}{2.5}\\	
\hline
Detector format	& \multicolumn{2}{|c|}{8k x 8k x 13 $\mu m$}& \multicolumn{2}{|c|}{2 x 8k x 8k x 13 $\mu m$}\\
\hline
Camera focal length, mm& \multicolumn{2}{|c|}{1700}& \multicolumn{2}{|c|}{1500}\\
\hline
Collim. focal length, mm& \multicolumn{2}{|c|}{1878}& \multicolumn{2}{|c|}{1633}\\
\hline
Beam diameter, mm	& \multicolumn{2}{|c|}{72}& \multicolumn{2}{|c|}{49}\\
\hline
Echelle frequency, $mm^{-1}$	&82.7& 202.43 & 237.2 & 474.3 \\
\hline
Echelle blazing angle,$^\circ$	&52.3&	42.9& 44.4 & 44.4 \\
\hline
Cross disp. frequency, $mm^{-1}$	& 267 & 539 & 318 & 637  \\
\hline
\end{tabular}
\end{center}
\end{table} 

\begin{figure} [!ht]
   \begin{center}
   \begin{tabular}{c} 
   \includegraphics[width=0.6\textwidth]{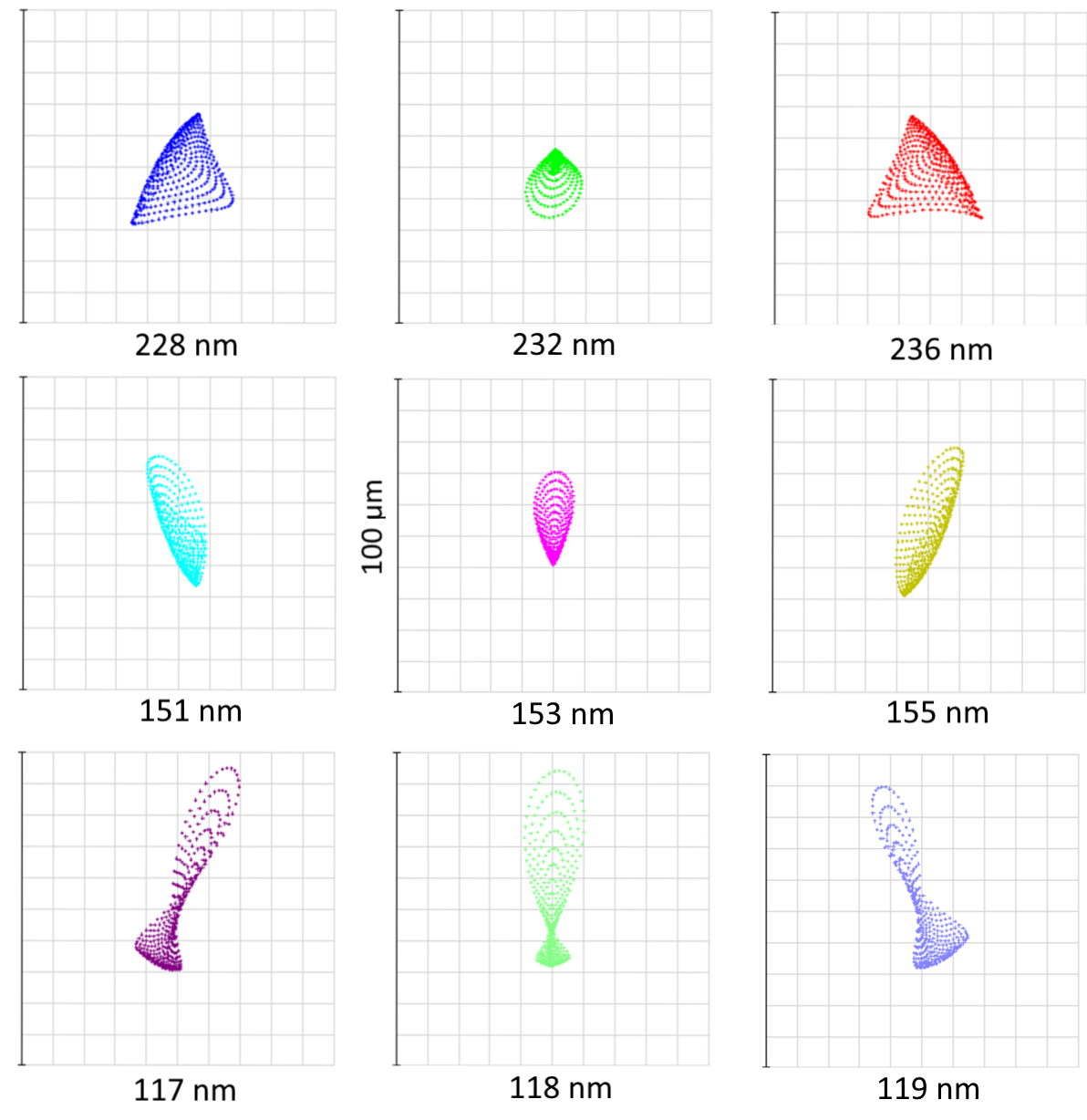}
   \end{tabular}
   \end{center}
   \caption[muvToroidSpots] 
   { \label{fig:muvToroidSpots} 
Spot diagrams of the MUV spectrograph for the 8m telescope. The cross-disperser is a toroidal concave grating.}
   \end{figure}

Figure~\ref{fig:muvToroidSpots} shows the spot diagrams for the MUV channel with the 8m telescope. Note that due to the grating equation scaling we can expect the NUV channel to be geometrically similar including the aberrations. In this particular case the necessary aberration correction may be reached just by using as slow toroid as the cross-disperser grating substrate. Its radius of curvature equals to   3394.02/3392.19 mm in the tangential and sagittal sections, respectively. The overall aberration distribution has an obvious XY symmetry. Another notable feature of the spot diagrams is a good aberrations correction in the sagittal directions, necessary for separation of the spectral orders and polarization components within each order. 

In this case the MUV, whose design is more challenging by default, reaches the required performance in terms of the simultaneous wavelengths coverage and spectral resolving power with a good safety margin - see Table~\ref{tab:core8} and Table~\ref{tab:core8nuv}. In addition, this solution leaves good margins for scaling and tuning of the designs (note the differences between the two channels in Table~\ref{tab:corePar} ). Therefore, it becomes possible to reach even a better performance in the NUV channel.

 \begin{table}[!ht]
\caption{Spectral resolving summary for the MUV channel with 8m telescope.} 
\label{tab:core8}
\begin{center}       
\begin{tabular}{|c|c|c|p{2.5cm}|} 
\hline
Order & Wavelength,nm & R & Full order length, nm 	\\	
\hline
57	&117.1	&94'046	&4.1	\\
\hline
	&118.1	&95'657	&	\\
 \hline
	&119.1	&97'268&			\\
 \hline
44	&151.3	&96'732	&5.3	\\
\hline
	&153.0	&95'645&			\\
 \hline
	&154.7	&94'558&		 \\	
 \hline
29	&228.2	&96'319	&8.0	\\
\hline
	&232.2	&95'481&		\\	
 \hline
	&236.2	&94'644&		 \\	
 \hline

\end{tabular}
\end{center}
\end{table} 

\begin{table}[!ht]
\caption{Spectral resolving summary for the NUV channel with 8m telescope.} 
\label{tab:core8nuv}
\begin{center}       
\begin{tabular}{|c|c|c|p{2.5cm}|} 
\hline
Order & Wavelength,nm & R & Full order length, nm	\\	
\hline
81	&234.7	&131'263	&5.8\\
\hline
	&236.1	&132'873	& \\
 \hline
	&237.5	&134'482&	\\
 \hline
62	&305.9	&133'918&	7.5\\
\hline
	&308.4& 132'846&	\\
 \hline
	&310.9	&131'775& \\	
 \hline
41	&460.7	&133'670&	11.4\\
\hline
	&466.4	&132'850& \\	
 \hline
	&472.1&	132'030& \\	
 \hline

\end{tabular}
\end{center}
\end{table} 

\begin{figure} [!ht]
   \begin{center}
   \begin{tabular}{c} 
   \includegraphics[width=0.9\textwidth]{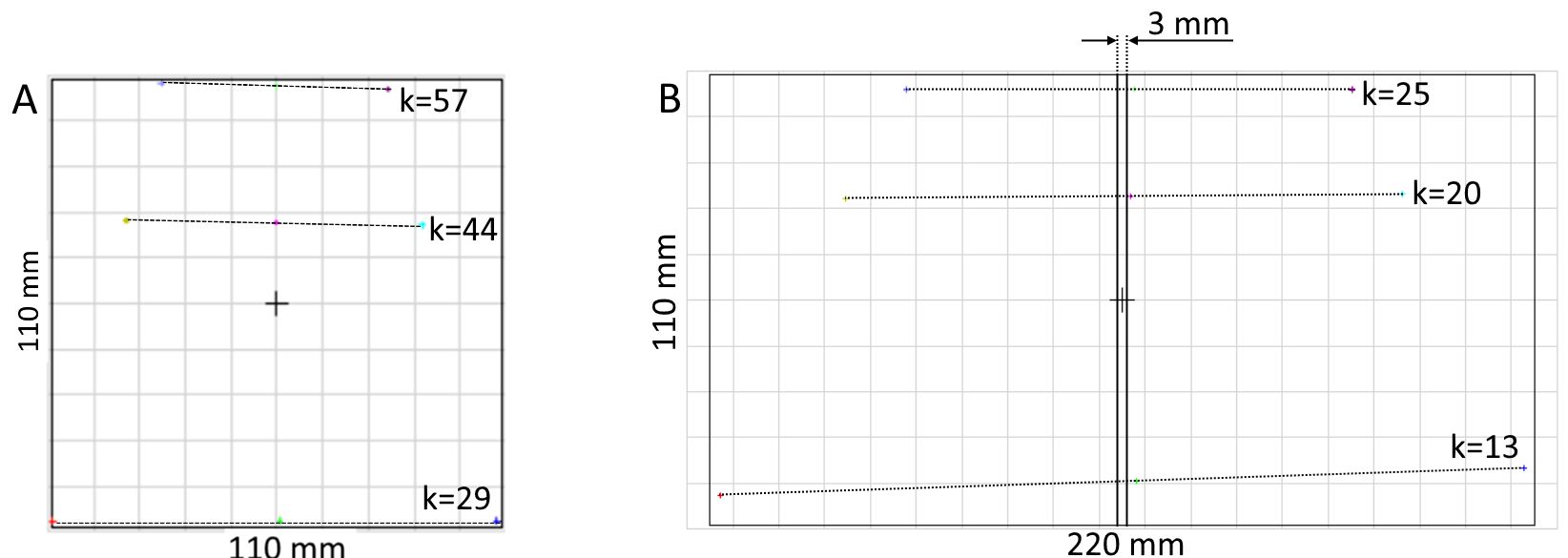}
   \end{tabular}
   \end{center}
   \caption[filling] 
   { \label{fig:filling} 
MUV channel 2D spectrogram filling diagram: A -- design for the 8m telescope with a monolithic detector, B -- design for the 6m telescope with detectors tiling.}
   \end{figure}

Once we switch to the reduced 6m primary diameter option, it becomes more difficult to reach the required performances. In turn, this leads us to some further trade-offs and compromise decisions. First of all, with a smaller telescope collecting area it becomes impossible to reach the required spectral resolving power and the order length at the same time while keeping the detector length equal to that of currently available sensors. Then we have to either assume that the production of a very large detector will become possible in a near future, or consider tiling of a few individual sensors. The first option is less probable as it would require a level of customization, barely possible in the semiconductors industry and/or a significant effort in the technological development. In the latter case we would have to deal with a blind zone in each order, but also increase the linear dispersion to keep the length of the spectrum fraction detected by each pixels' row above the specified value. The MUV channel detector filling, including the blind zone effect, is illustrated by Fig.~\ref{fig:filling}. We assume that the minimum gap between the 8k class detectors will be around 3 mm, as it was shown for some other instruments \cite{Thatte2022SPIE12184E..20T}. The spectral width of the corresponding blind zone is given in the summary Table~\ref{tab:core6}. On the other hand, using a larger detector allows to work in lower orders and decrease their total number. As the two detector units have identical Y dimensions, this would mean a better separation of the orders and polarization components.

\begin{figure} [!ht]
   \begin{center}
   \begin{tabular}{c} 
   \includegraphics[width=0.75\textwidth]{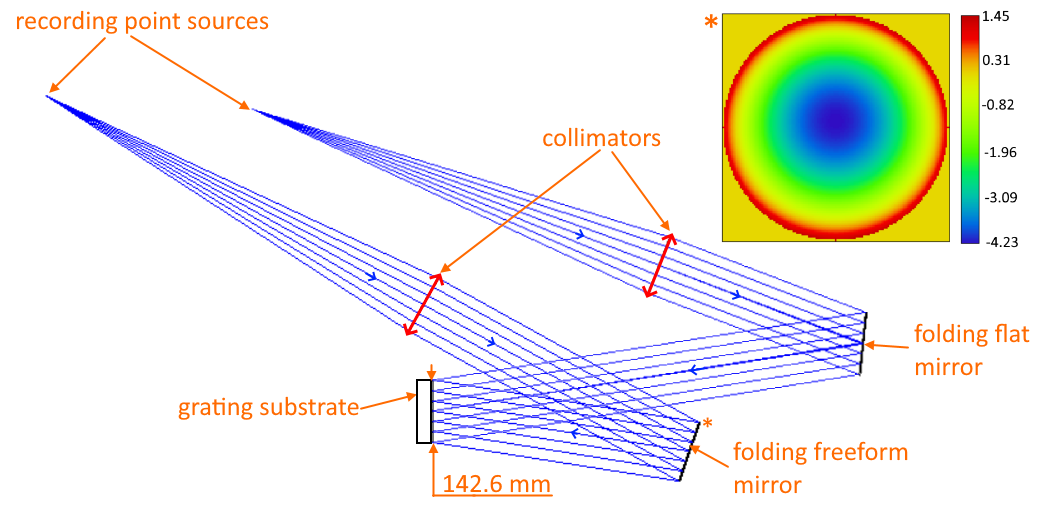}
   \end{tabular}
   \end{center}
   \caption[recording MUV] 
   { \label{fig:recording MUV} 
Recording setup for the MUV cross disperser grating and profile of auxiliary deformable mirror (sag deviation in microns).}
   \end{figure}

Another effect, which we have to face when adapting the design for a smaller telescope is the aberration correction. For the aforementioned reasons we have to increase the dispersions of both of the gratings. This means growing angles on incidence and exitance for the cross-disperser grating. One solution to add more correction parameters into the design without using additional surfaces is a holographic recording with aberrated wavefronts. This approach implies using some auxiliary optics in a recording setup used to form the necessary grooves pattern and such an element is called in some cases a $2^{nd}$ generation holographic grating \cite{Palmer:89}. The blazed grating profile may be formed afterwards by some etching technology, or formed directly by inversion of one of the recording wavefronts. In Figure~\ref{fig:recording MUV} we show a variant of recording setup layout for the MUV grating to be used in the 6m instrument version. One of the recording beams is deformed by a deformable mirror. As the diagram shows, the required sag deviation is small, though this component should be of a large diameter. Alternatively, it can be moved into a diverging beam before the collimator. In this particular case we limited the surface shape definition of this auxiliary mirror by the YZ-symmetrical first-order Zernike modes. However, with currently available deformable mirrors the deformation pattern can be more complex.

\begin{figure} [!ht]
   \begin{center}
   \begin{tabular}{c} 
   \includegraphics[width=0.6\textwidth]{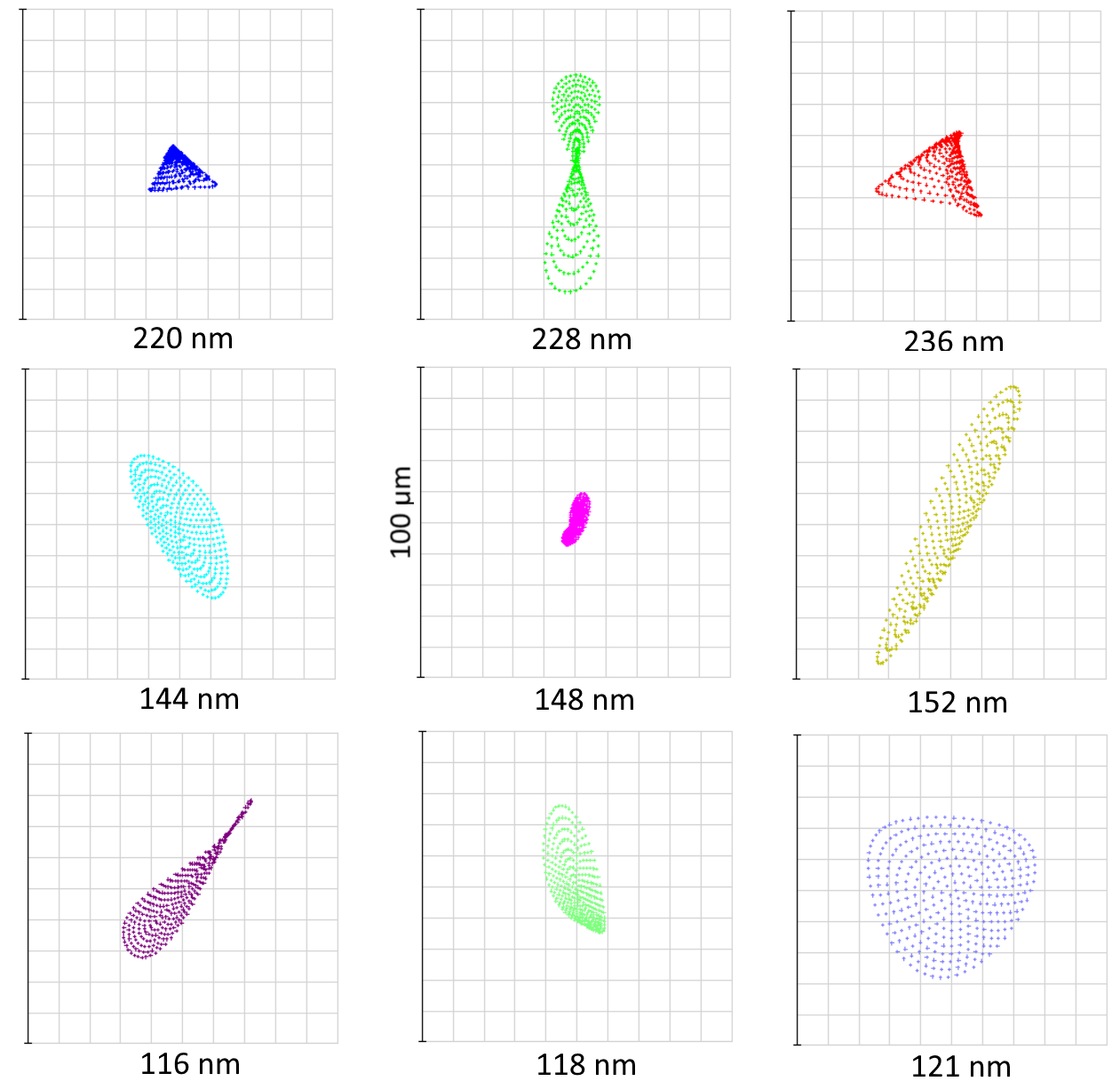}
   \end{tabular}
   \end{center}
   \caption[muvOFHSpots] 
   { \label{fig:muvOFHSpots} 
Spot diagrams of the MUV spectrograph for the 6m telescope. The cross-disperser is a holographic concave grating, recorded with a freeform wavefront.}
   \end{figure}

To demonstrate the aberrations changes caused by the dispersion increase, we plot the spot diagrams in Fig.~\ref{fig:muvOFHSpots}. This case corresponds to the corrected holographic grating. It shows that we have to deviate from the double-plane symmetry when using the freeform recording wavefront. Also it indicates a compromise between the correction in main- and cross-dispersion directions. 

\begin{figure} [!ht]
   \begin{center}
   \begin{tabular}{c} 
   \includegraphics[width=0.75\textwidth]{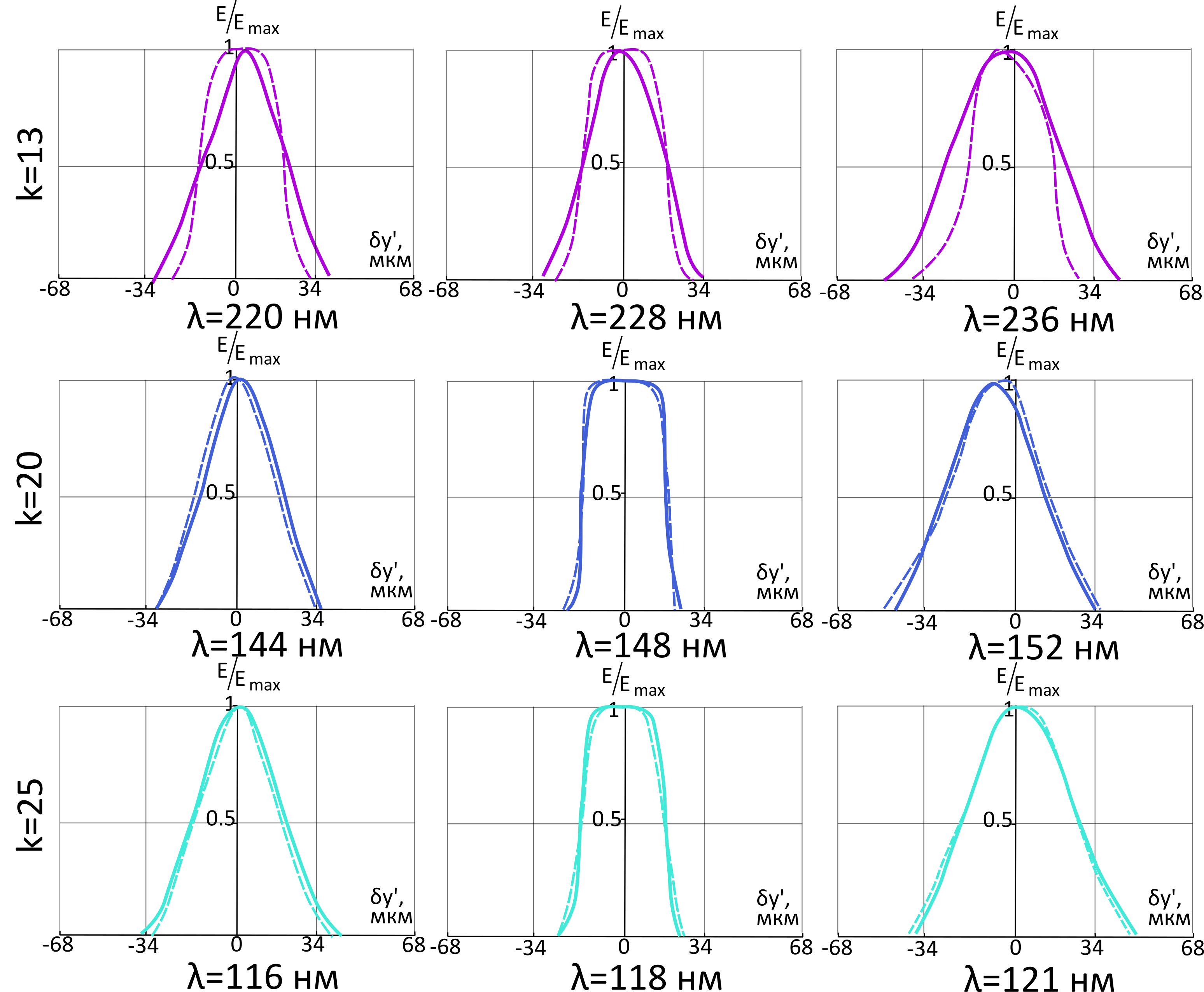}
   \end{tabular}
   \end{center}
   \caption[IFs] 
   { \label{fig:IFs} 
Instrument function of the spectrograph for the 6m telescope: solid curve -- toroidal cross-disperser, dashed curve -- $2^{nd}$  generation holographic cross-disperser.}
   \end{figure}

   Separately, we analyze the correction effect introduced by the recording setup shown in Figure~\ref{fig:recording MUV}. In Fig.~\ref{fig:IFs} we compare the instrument functions, i.e. relative irradiation distribution in a monochromatic image of the entrance pinhole. The solid lines correspond to a toroidal cross-disperser, equivalent to the one used before in the 8m version, and the dashed lines correspond to the corrected holographic grating.  The plots clearly show that the additional correction parameters allow us to get a better resolution in the longest order, i.e. at the longwave side of the spectrum. At the same time the effect for the rest of the spectrogram is moderate.  

The numerical values of the resolving power and the orders spectral length for both versions of the 6m  MUV design are summarized in Table~\ref{tab:core6}. In general we can conclude that decreasing the telescope aperture would require more complex solutions in the spectrographs' design to maintain the key performance metrics. This may mean lower technology readiness, so a higher cost and longer development time. However, the resolution and spectral coverage targets remain reachable under certain conditions.

   \begin{table}[!ht]
\caption{Spectral resolving summary for the MUV channel versions for the 6m telescope.} 
\label{tab:core6}
\begin{center}       
\begin{tabular}{|l|p{2cm}|p{2cm}|p{2cm}|p{2cm}|p{2cm}|} 
\hline
Order &	Wavelength, nm	& R with toroidal cross-disp.&	R with holographic cross-disp. &	Tiling gap, nm	& Order length, nm\\
\hline
25	& 116.1	& 87'628&	88'437&	0.2&	9.1\\
\hline
	&118.5&	91'751&	91'751	&    	& \\ 
 \hline
	&120.9&	67'458&	95'064	&	    &   \\
 \hline
20	&144.4	&77'359&	77'359&	0.1	&11.4\\
\hline
	&148.1	&91'728&	91'728&		&  \\
 \hline
	&151.8	&89'435&	89'435&    &  \\		
 \hline
13	&219.3	&66'290&	67'921&	0.1	&17.5\\
\hline
	&228.1&	91'586&	91'586&	&	\\
 \hline
	&236.9	&82'635&	85'044& &	\\	
 \hline

\end{tabular}
\end{center}
\end{table} 

\begin{figure} [!ht]
   \begin{center}
   \begin{tabular}{c} 
   \includegraphics[width=0.7\textwidth]{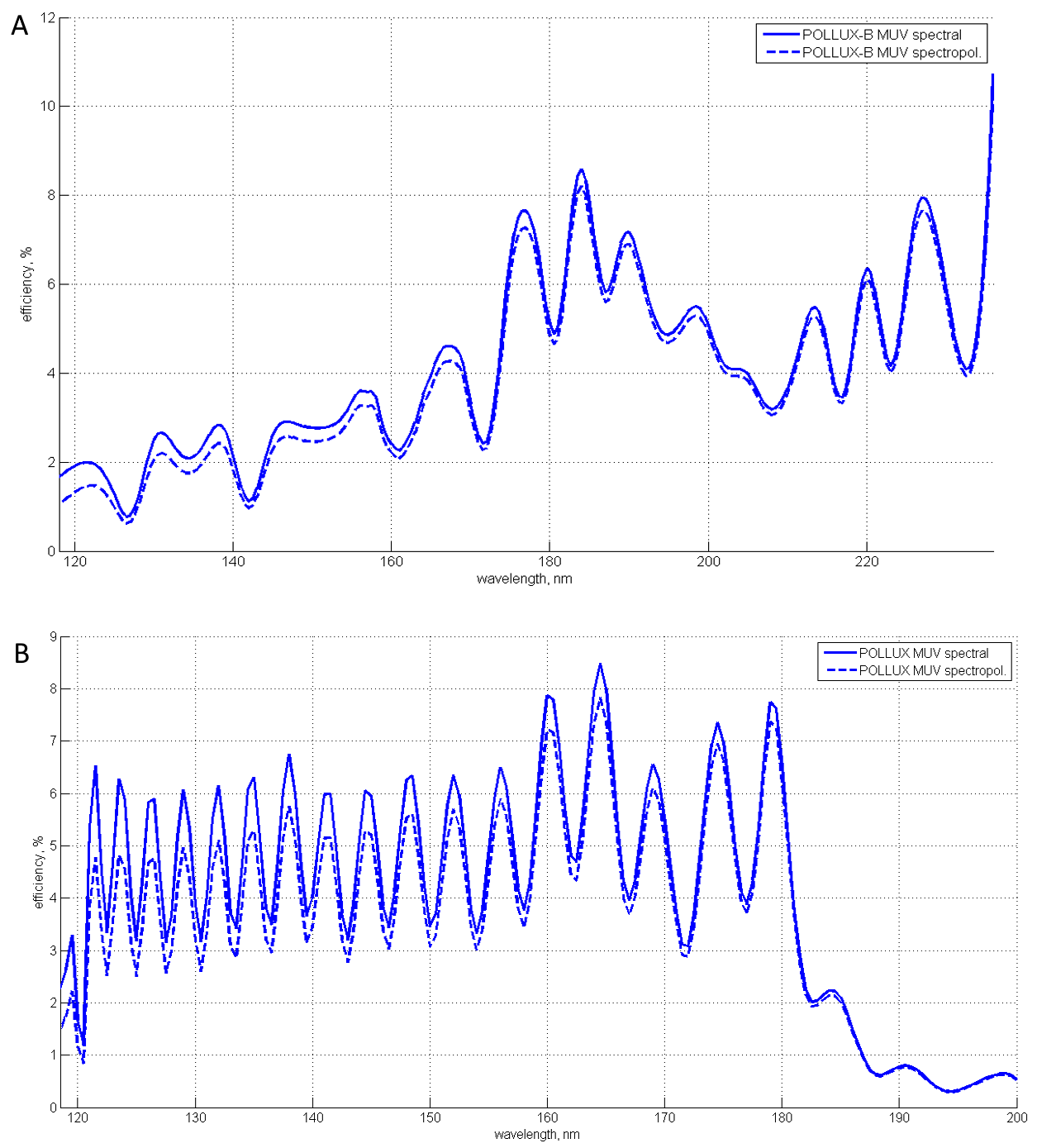}
   \end{tabular}
   \end{center}
   \caption[effic] 
   { \label{fig:effic} 
Overall efficiency of the MUV channel including the optical train throughput and the EMCCD QE: A -- current design for the 8m telescope, B -- design from the LUVOIR proposal.}
   \end{figure}

As a separate point we consider the end-to-end efficiency of the MUV channel. We take the less-risky 8m version of the design and compute the overall optical train throughput. We compute the echelle efficiency for the specified parameters using the GSolver software\cite{Johnson}. Also, we apply the latest reflectivity curve for a UV-enhanced eLiF coating, known from the literature\cite{Quijada2022SPIE12188E..1VQ} as we expect that such a coating will be available in the nearest future and applied for the entire HWO telescope (see also paper by T. M. Glassman at this conference). For the rest of the components we assume the same throughput as we had in the previous Pollux study\cite{Muslimov10.1117/12.2310133}. We also include the EMCCD quantum efficiency\cite{Nikzad_s16060927} in the budget. The results are shown in Fig.\ref{fig:effic}, A. For comparison we plot a similar transmission curve calculated before for the Pollux at LUVOIR-A project (Fig.\ref{fig:effic}, B). In general, the two curves are approximately at the same level. The maximum efficiency is defined mainly by the reflective coating properties, so the new design exhibits a better performance at the longwave end. For the same reason it has a lower efficiency towards the Lyman $\alpha$ end. Change of the orders number and dispersion leads to a slight change in the efficiency variation within each diffraction order. All of these effects are minor in comparison with the effective area change defined by the hosting telescope parameters.    

\section{Visible to Near infrared: defined by the detector and science cases}
\label{sec:visnir}  

High-resolution spectroscopy and spectropolarimetry at visible and near infrared wavelengths could increase the scientific value of the proposed instrument drastically. This is suggested by some publications on the topic \cite{Birkby2018arXiv180604617B} and by direct feedback from the science groups during and after the decadal survey activities. We leave the exact list of science cases, which would require this mode, for a separate publication and focus here only on the corresponding technical limitations.

To the best of our understanding, the available detector options may be the main limiting factor for the composition of such a VIS-NIR channel. Typical silicon-based CCD and CMOS sensors hardly allow to reach the maximum wavelength of 1050 nm \cite{CCD}. Even with a specific doping, they reach 1100 nm\cite{Blaze}, which is not enough to cover the main range of interest. However, there is a variety of commercial solutions of this type certified for space application and the lower spectral limit may be set to match the end of the NUV sub-range. 

\begin{figure} [!ht]
   \begin{center}
   \begin{tabular}{c} 
   \includegraphics[width=0.9\textwidth]{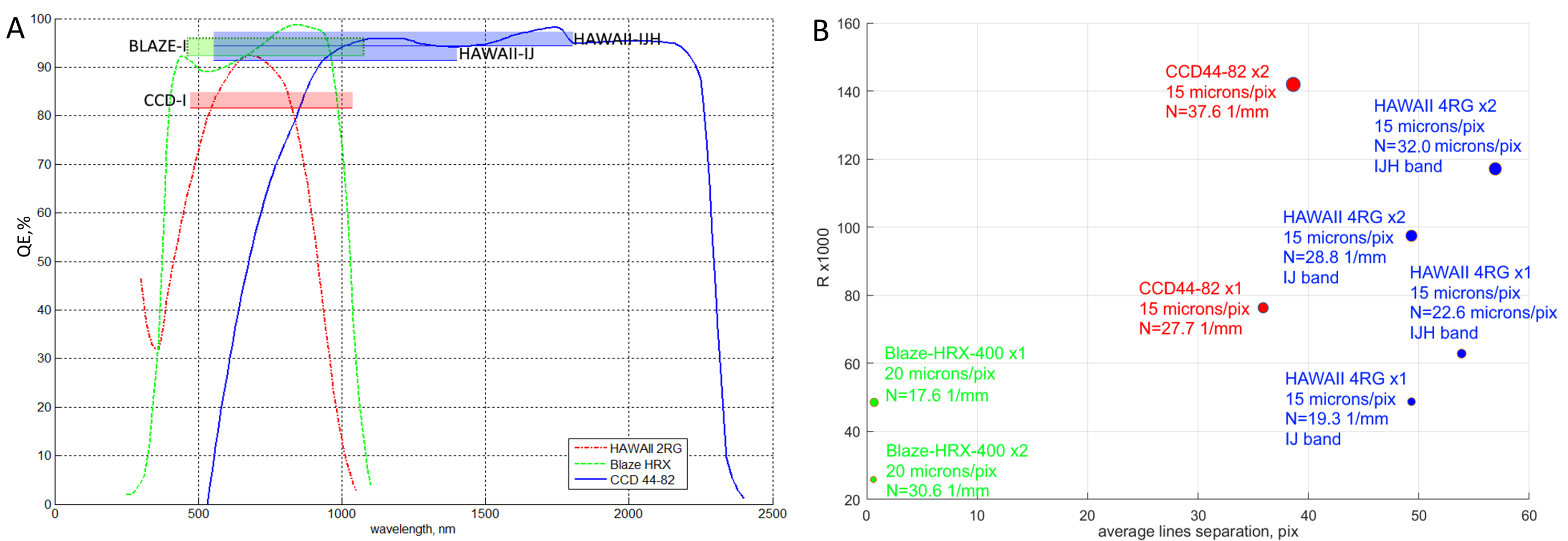}
   \end{tabular}
   \end{center}
   \caption[visDet] 
   { \label{fig:visDet_tr} 
VIS-NIR channel detector trade-off inputs: A -- quantum efficiencies of typical detectors, B -- spectral image line separation VS spectral resolving power for a few commercial detectors.}
   \end{figure}

An alternative solution could be based on a \textit{CdHgTe} sensor\cite{Jerram10.1117/12.2536040} such as HAWAII, used for the HARMONI instrument\cite{Thatte2022SPIE12184E..20T}. Its application for a space instrument should not cause any difficulties, but the selection of the working spectral band becomes not that straightforward. First of all, the quantum efficiency curve (see Fig.~\ref{fig:visDet_tr},A) indicates that the lower limit cannot reach below 560 nm. Second, if we decide to use such a detector, the upper limit could be extended towards 1800 nm to cover the I,J and H bands simultaneously. However, such a solution would require using a large number of working diffraction orders, which in turn would create issues with the orders and components separation, aberration correction and technical feasibility of the cross-disperser. These difficulties could be mitigated by using the detector tiling, but it would create blind zones as discussed in Sec.~\ref{sec:muvnuv}. Finally, if we try to follow the parameters of commercial detectors as their pixels sizes, number of pixels and aspect ratio, the reachable spectral resolving power can be also limited. All of these trade-offs are summarized in Fig.~\ref{fig:visDet_tr},B. 

\begin{figure} [!ht]
   \begin{center}
   \begin{tabular}{c} 
   \includegraphics[width=0.7\textwidth]{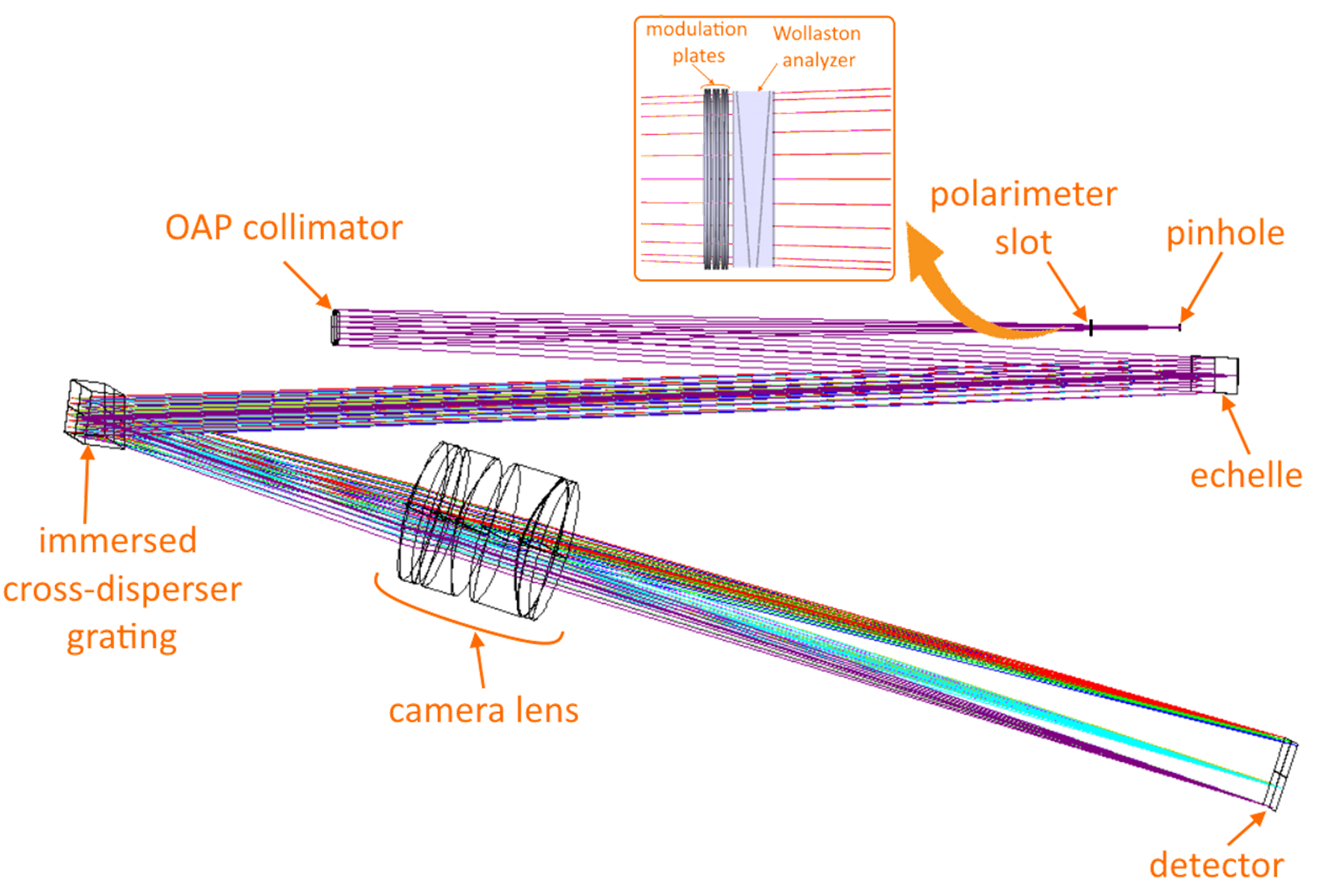}
   \end{tabular}
   \end{center}
   \caption[vis] 
   { \label{fig:visDet} 
General view of the VIS-NIR channel optical design.}
   \end{figure}

We are not pursuing the goal to cover all of the mentioned design options here.  Instead we demonstrate  an implementation for one of them, which is associated with the lowest technological risks - an echelle spectrograph for the 8m telescope based on a single silicon CCD sensor. This decision limits the working spectral range to 472-1050 nm. In general, the architecture of such a VIS-NIR channel may be quite similar to the one presented in Fig.~\ref{fig:muv}. The polarimetric unit can use the same basic design, but with material adapted to the wavelength as used in ground-based spectropolarimeters\cite{Donati2003ASPC..307...41D}. The collimator and the main disperser, i.e. the echelle grating will be approximately the same. However, we expect that the simultaneously detected spectral range will be larger, so it is unlikely that the required orders separation and aberrations corrections over a wide field and spectral band could be reached with a simple single-element cross disperser and camera. Instead, we  propose to use an immersive grating, which was successfully applied in some other instruments \cite{Amerongen10.1117/12.925612}, and a separate refractive camera lens. Fortunately, for a limited spectral range and dispersion it becomes possible to reach the required image quality with a simple all-spherical Tessar-type design. The design obtained with these solutions is shown in Fig.~\ref{fig:visDet}. 

\begin{figure} [!ht]
   \begin{center}
   \begin{tabular}{c} 
   \includegraphics[width=0.6\textwidth]{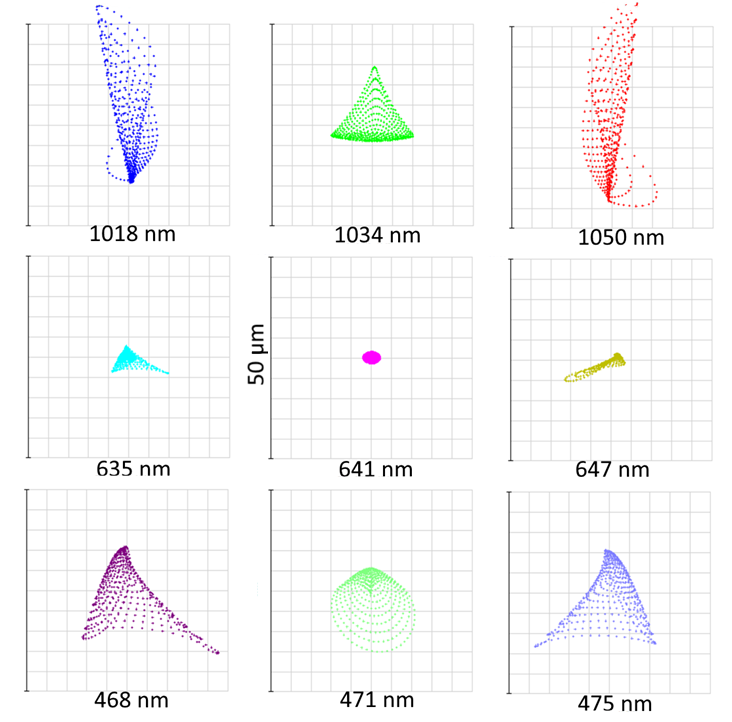}
   \end{tabular}
   \end{center}
   \caption[visSpots] 
   { \label{fig:visSpots} 
Spot diagrams of the VIS-NIR spectrograph for the 8m telescope based on an ordinary CCD.}
   \end{figure}

To give an impression of the image quality reached in this case, we provide the spot diagrams  (see Fig.~\ref{fig:visSpots}). The aberration pattern over the field is relatively symmetrical, however, they notably change across the spectrum. It can be noted that the astigmatic elongation of the spot in the cross-dispersion direction is sacrificed to get a better spectral resolution. However, usage of this approach is restricted by the orders separation, which may become only tighter as we try to work in a wider band. Also, one can see that even if the spectrogram centre is perfectly corrected, the aberrations grow towards its corners rapidly. Similarly, this effect will restrict the channel performance if we try to extend the simultaneously detected spectral range. Using aspheres and catadioptric systems can help to correct the aberrations, but this would make the camera design more complex and sensitive. 

The main design parameters for this version of the VIS-NIR channel are summarized in table~\ref{tab:visPar}.

   \begin{table}[!ht]
\caption{Key parameters of the VIS-NIR channel for the 8m telescope based on an ordinary CCD.} 
\label{tab:visPar}
\begin{center}       
\begin{tabular}{|c|c|} 
\hline
\multicolumn{2}{|c|}{Collimator}\\
\hline
Diameter, mm &	48 \\
\hline
Focal length, mm	& 1200 \\
\hline
Off-axis shift, mm	& 64 \\
\hline
\multicolumn{2}{|c|}{Echelle}\\
\hline
Frequency, $mm^{-1}$ &	48 \\
\hline
Blazing angle, $^{\circ}$	& 49.6\\
\hline
Size, mm &	48 x 72 \\
\hline
\multicolumn{2}{|c|}{Cross-disperser}\\
\hline
Prism &	7.59$^{\circ}$ F.Silica wedge \\
\hline
Frequency, $mm^{-1}$ &	 120 \\
\hline
Size, mm&	160x72 \\
\hline
\multicolumn{2}{|c|}{Camera}\\
\hline
Focal length, mm&	1253 \\
\hline
Max diam., mm&	184 \\ 
\hline
Type&	All-spherical Tessar\\
\hline
\multicolumn{2}{|c|}{Performance}\\
\hline
Order length, nm&	15.1-33.4 \\
\hline
R&	 $\leq77'042$\\ 
\hline
Sampling, pix&	2.55\\
\hline

\end{tabular}
\end{center}
\end{table} 
 To estimate the reachable performance metrics an interested reader may address those of X-Shooter instrument\cite{Vernet2011A&A...536A.105V}, which inspired this design to a large extent. In contrast with the UV channels, we do not expect major difficulties with the throughput in this case, although design of  broadband coatings and assuring their durability in the space environment may become a separate challenge. Generally, all of the components are technologically feasible with a minimum risk and the performance matches the requirements with some safety margins. However, as we mentioned before, this solution may not be scalable for a wider or red-shifted working band. 
 
\section{Far UV: compromise between size, resolution and throughput}
\label{sec:FUV}  

Observations in the far UV is a unique capability of space instruments. Unfortunately, there were few space missions since the previous generation instruments as STIS\cite{Woodgate_1998}, GALEX\cite{Milliard10.1007/10849171_18} and COS\cite{Green_2012} providing UV spectroscopy below 200 nm, while the spectropolarimetric measurements in that domain remain unprecedented. Moreover, high resolution spectroscopy and spectropolarimetry deeper in the far UV (FUV) domain have not been implemented so far. Therefore, adding this functionality to the proposed instrument seems to be an attractive asset. As we will have a large collecting area and a UV-optimized coating, even a moderate efficiency could be enough to get valuable scientific data, which is impossible to get in any other way. 

Performing measurements below the lower limit of MUV and reaching at least 100 nm would require a dedicated channel. The only feasible solution for polarimetry in this range is based on a modulator consisting of three grazing-incidence mirrors and a Brewster angle analyzer. The technological readiness of this solution is growing\cite{LeGal:20} (also see Girardot et al., at this meeting) and we hope to make it suitable for a flagship space mission in the near future. 

\begin{figure} [!ht]
   \begin{center}
   \begin{tabular}{c} 
   \includegraphics[width=0.6\textwidth]{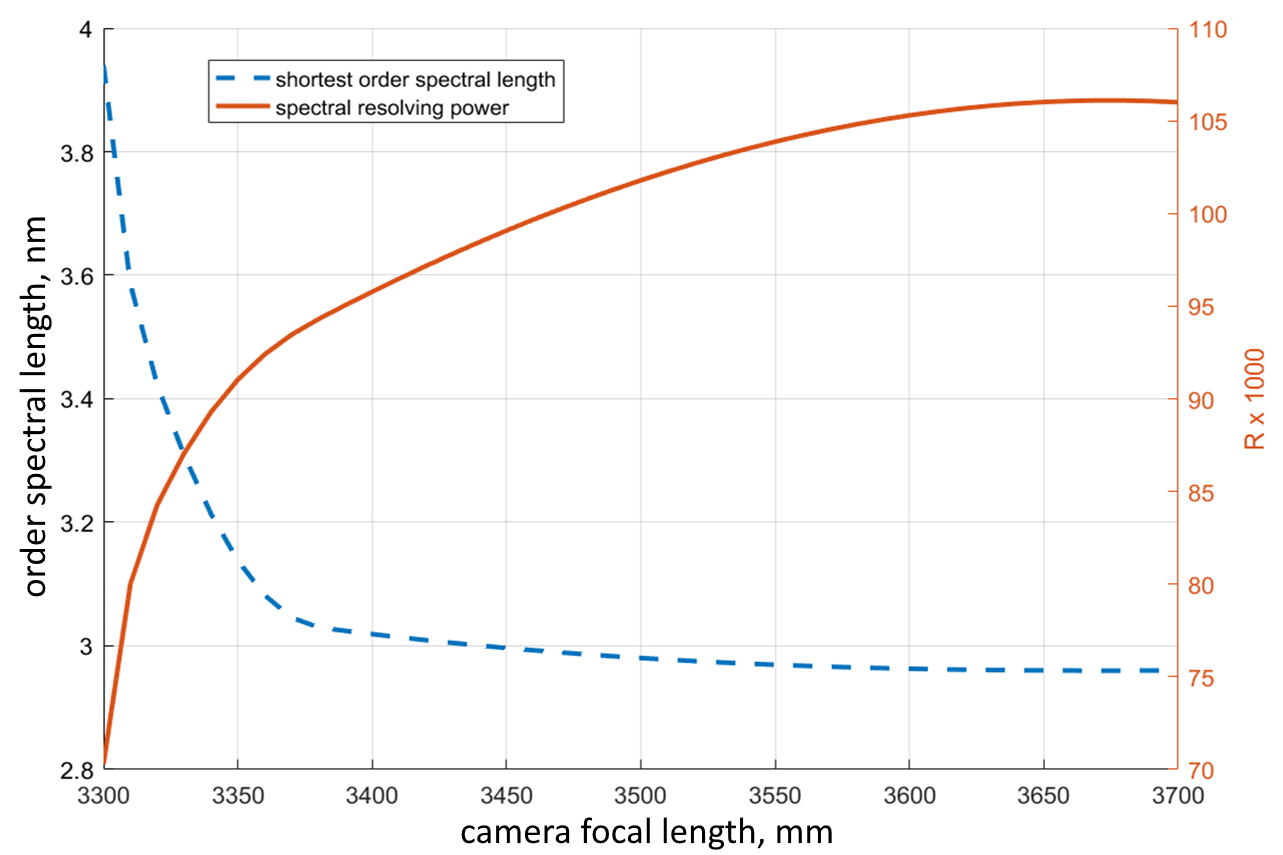}
   \end{tabular}
   \end{center}
   \caption[fuvLength] 
   { \label{fig:fuvLength} 
Fundamental limitations for the resolving power and free spectral interval of the FUV echelle spectrograph.}
   \end{figure}

\begin{figure} [!ht]
   \begin{center}
   \begin{tabular}{c} 
   \includegraphics[width=0.75\textwidth]{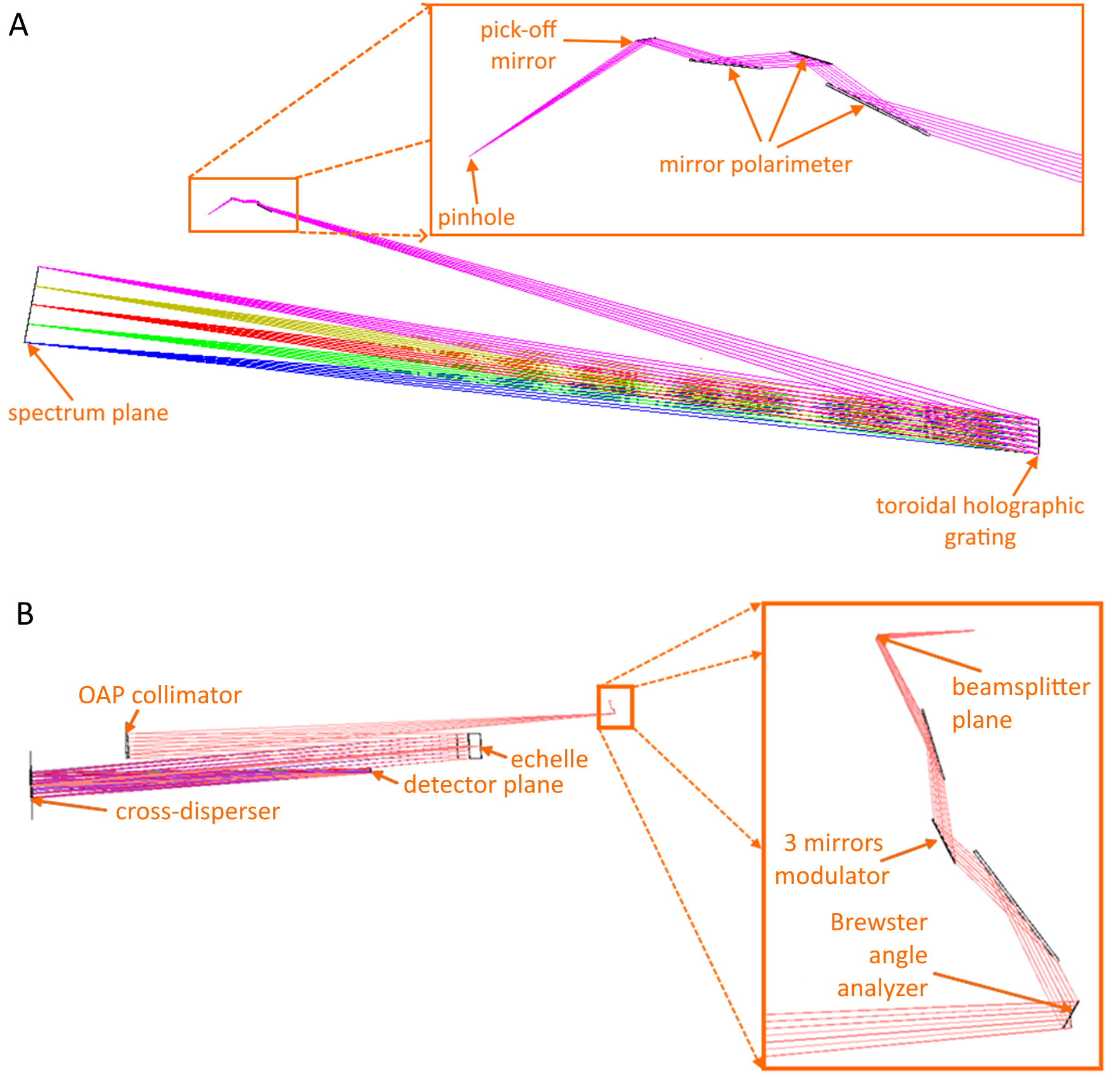}
   \end{tabular}
   \end{center}
   \caption[fuv] 
   { \label{fig:fuv} 
FUV channel optical design options: A -- spectrograph with a single concave grating, B -- echelle-spectrograph.}
   \end{figure}

The target high resolving power would also imply using an echelle spectrograph. However, the equations defining the echelle geometry indicate that the target specifications may lead to very challenging parameters combinations.  Fig.~\ref{fig:fuvLength} illustrates the relationship between such key performance indicators as the spectral resolving power and the spectral order length and the camera focal length, which defines the instrument size. If we try to get a high dispersion and limit  the spectral length of the orders and their number, we should increase the detector length and the linear dispersion. The latter parameter depends on the echelle spatial frequency, which has strict technological limitations, and the camera focal length. All of these factors together lead to solutions based on long focal length optics. Moreover, due to the basic equations non-linearity, the effect of focal length increase, is limited as the curves visually show. This simple estimation  together with the trade-offs shown in Sec.~\ref{sec:muvnuv} indicate that with a 6m telescope the FUV channel may become unreasonably large and complex. 
However, for the 8m telescope we can propose a solution, based on a certain mitigation of the spectral order length. It is very close to that proposed before for LUVOIR\cite{Muslimov10.1117/12.2310133}. Its general view is shown in Fig.~\ref{fig:fuv}, B and the corresponding key parameters are shown in table~\ref{tab:fuvPar}. 

Despite the feasibility of the key components in this echelle spectrograph design, we would like to consider an alternative, namely a spectrograph with single concave grating. This approach would be an obvious solution for the low reflectivity of known coatings in the FUV and the technological difficulties of large high-frequency echelle gratings production. However, the spectral resolving power in this case will be limited to a few tens of thousands, which is at the very edge of the domain traditionally referred to as a "high resolution". As a consequence, this may make some of the science goals in the FUV unreachable. This design is also obviously dependent on the instrument size.

\begin{figure} [!ht]
   \begin{center}
   \begin{tabular}{c} 
   \includegraphics[width=0.6\textwidth]{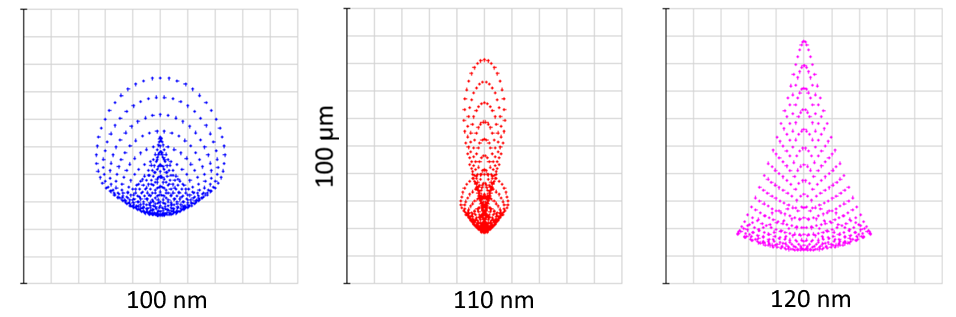}
   \end{tabular}
   \end{center}
   \caption[fuvSpots] 
   { \label{fig:fuvSpots} 
Spot diagrams of the FUV spectrograph for the 8m telescope based on a single concave grating.}
   \end{figure}

   The corresponding optical design is shown in Fig.~\ref{fig:fuv}, A. Our estimations show that the concave grating can be recorded holographically at 488 nm on a spherical surface with two beams, one of which is parallel with $+75^\circ$ angle incidence, and the second one is defocused by	3047 mm and has the angle of $-60.5^\circ$.  The corresponding spot diagrams are shown in Fig. ~\ref{fig:fuvSpots}.

The main parameters of the optical components and the performance metrics for both of the design alternatives are given in Table~\ref{tab:fuvPar}. 

\begin{table}[!ht]
\caption{Key parameters of the FUV channel options for the 8m telescope.} 
\label{tab:fuvPar}
\begin{center}       
\begin{tabular}{|c|c|c|} 
\hline
Parameter	&Single grating& Echelle spectrograph\\
\hline
Min wavelength, nm&	100 &	100 \\
\hline
Max wavelength, nm&	120 &	120 \\
\hline
Sampling, pix &3.3 &	2.55 \\
\hline
Detector format	&10k x 3k x13$\mu m$	& 8k x 3k x13 $\mu m$\\
\hline
Camera focal length, mm	&1640 &	1700\\
\hline
Collim. focal length, mm	& - &	1878\\
\hline
Beam diam.,mm& - &	 72.2 \\
\hline
Main disp. frequency , $mm^{-1}$&	3700&	348\\
\hline
Blazing angle, $^\circ$&	7.25&	49.65\\
\hline
Orders&	1	&37-44\\
\hline
Order length, nm	&13.0&	2.7-3.2\\
\hline
R	& $\leq18'700$&	$\leq 120'750$\\

\hline

\end{tabular}
\end{center}
\end{table}

\section{Discussion}
\label{sec:disc}  
In the present study we considered and compared a few optical design options for the Pollux spectrolarimeter, which can become the European contribution for the future Habitable Worlds Observatory. We do not claim that it is comprehensive and covers all the possible options and the corresponding engineering trade-offs. However, we hope that it will be useful for the science groups and system engineers for estimation of the reachable parameters or ranking and prioritizing different options. 

In general, we can conclude that the MUV and NUV channels for an 8m telescope can be built based on low-risk technologies and components and meet the required performance. If the primary diameter is reduced to 6m, the solution still can be found, though it may be less favourable in terms of the performance and technological readiness. 

The option of additional VIS-NIR channel has wide prospects in terms of the science goals and is relatively safe in terms of the technologies used. The baseline design choice will be driven mainly by the  detectors available.  

The FUV channel  may be the most challenging option. An echelle spectrograph for this region could be built with a certain mitigation of the resolution and orders stitching requirement. An alternative single-grating design would require a higher throughput and a simpler manufacturing and alignment in expense of notable downgrade in spectral resolution.  

There are a few important questions left outside of the scope of this paper. It is worth at least to mention the sensitivities of the proposed optical designs to the manufacturing and alignment errors, wavefront error budget compilation, impact of the telescope architecture on the polarimetric measurements precision, stability of the instrument together with the telescope and the spacecraft. Each of this points may change our estimates of the design solution feasibility. However, each of them should be thoroughly analyzed in a separate work.   

\bibliography{main}
\bibliographystyle{ieeetr}

\end{document}